# The Impact of Downgrading Protected Areas (PAD) on Biodiversity

Authors

Yufei Li
School of Economics and Management, Beihang University, Beijing, China

Lingling Hou
School of Advanced Agricultural Sciences, Peking University, Beijing, China

Pengfei Liu
Department of Environmental and Natural Resources Economics, University of Rhode Island, Kingston, RI 02881, United States, Email: pengfei_liu@uri.edu, Corresponding author.

**Abstract**
We quantitatively assess the impacts of Downgrading Protected Areas (PAD) on biodiversity in the U.S.. Results show that PAD events significantly reduce biodiversity. The proximity to PAD events decreases the biodiversity by 26.0% within 50 km compared with records of species further away from the PAD events. We observe an overall 32.3% decrease in abundance after those nearest PAD events are enacted. Abundance declines more in organisms living in contact with water and non-mammals. Species abundance is more sensitive to the negative impacts in areas where PAD events were later reversed, as well as in areas close to protected areas belonging to the International Union for Conservation of Nature (IUCN) category. The enacted PAD events between the period 1903 to 2018 in the U.S. lead to economic losses of approximately $689.95 million due to decrease in abundance. Our results contribute to the understanding on the impact of environmental interventions such as PAD events on biodiversity change and provide important implications on biodiversity conservation policies.

**JEL** Q13, Q48, Q51, Q57

# Introduction

The global population expansion increases the complexity of land use planning[1]. Protected areas (PAs) are geographical areas that have been specifically identified, declared, devoted, and administered by law or other regulations to preserve nature including the ecosystem services and cultural values it supports. Protected area downgrading, downsizing, and degazettement (PADDD) events represent legal modifications that weaken, shrink, or eliminate PAs, which can accelerate forest loss, fragmentation, and carbon emissions. In the continental U.S., protected area downgrading (PAD) events will lower the permitted volume, scope, or range of human activities within a protected area and affect biodiversity (see Figure 1).

Human intervention has led to adverse impacts and losses to nature and human systems. Growing demand for infrastructure, subsistence, industrial agriculture, minerals, and political pressure impose threats to PAs and the areas around PAs, which transform once-protected landscapes, threatening their biodiversity conservation value and associated ecosystem services[2]. While many studies have focused on the impacts of multiple types of land conservation expansion on biodiversity[1,3,4], there is no systematic research to quantify the threat of PAD events on biodiversity. This study quantitatively evaluates the impacts of PAD events on biodiversity with nationally representative, micro-level data in the U.S..

Broad legal changes often undermine the viability and effectiveness of protected areas[5]. Growing economic demand for minerals and other resources, as well as political pressure for related infrastructure such as road accessibility, creates new threats to biodiversity in protected areas. PAs are acknowledged in the Convention on Biological Diversity and the 2030 Agenda for Sustainable Development and regarded as a pillar of biodiversity conservation[6]. The long-term goal of area-based conservation needs to be supported by policy changes. PAD events may cause a variety of risks for local ecosystems, climate, and human society. Clear, transparent tracking of PAD events will ensure we correctly address current shortfalls in area-based conservation to contribute to biodiversity conservation related policies.

Despite requests to speed up the creation of protected areas to safeguard biodiversity, some governments have begun to pull down legal protections[7]. PADDD can reallocate under-performing protected areas (PAs), reducing PAs in regions that have limited development potential (such as remote areas, or those with steep slopes or high elevation)[8], while developing the technologies and infrastructure required for renewable energy production[9]. The first contemporary protected areas, Yellowstone and Yosemite National Parks, are located in the U.S., which has long been an example of global conservation. However, the U.S. government enacted at least 220 PADDD events (including 211 PAD events) between 1905 and 2018 in 195 terrestrial PAs in 46 states, repealing protections for a total of 22879.32 km$^2$. The first PADDD event occurred in Yosemite National Park when, in 1905, Yosemite was reduced in size by 30% to allow for forestry and mining. The cause of the majority of U.S. PADDD events (n = 186) was a 2016 National Park Service regulation enabling Native American tribes to harvest plants for customary subsistence purposes if the action will have "no significant ecological impact"[10]. Also, 33 PADDD events were connected to industrial-scale resource exploitation and development, including the downgrading of eight national forests to make room for the expansion of ski infrastructure in 1986.

The U.S. government has proposed more than 700 PADDD events that, if enacted, would affect hundreds of thousands of square kilometers of protected areas. Since 2000, 90% of U.S. PADDD proposals have been introduced by the government, and industrial-scale development has been involved in 99% of the proposals. The recent PADDD events highlighted the increasingly uncertain future of protected areas and biodiversity in the U.S. After 114 unsuccessful requests over 30 years, oil and gas drilling in the Arctic National Wildlife Refuge was authorized by the US Congress in 2017[11]. Some additional national monuments that favor biodiversity conservation have been downgraded by the U.S. government[12]. Since PAD events have the closest spatial impact on biodiversity in the continental U.S., our analysis focuses on PAD events.

We combine the BioTIME and PADDD tracker data to depict high-resolution distributions of organisms and PAD events across the U.S.. We then use a generalized difference-in-differences (DID) model to identify the impacts of PAD events on biodiversity. We show that PAD events have significant negative effects on biodiversity. The proximity to PAD events affects biodiversity negatively and decreases the abundance by 26.0% within 50 km. We also observe an overall 32.3% decrease in abundance after the nearest PAD event is enacted. Abundance declines more in organisms living in contact with water and non-mammals. Species abundance is more sensitive to the negative impacts in areas where PAD events were later reversed, as well as in areas close to protected areas belonging to the International Union for Conservation of Nature (IUCN) category. Our results offer valuable insights and benchmark statistics for policymakers to balance biodiversity protection and PAD in the U.S. and other countries.

An improved understanding of biodiversity change is important for future policymaking to deploy biodiversity conservation. We expect biodiversity change that threatens the resilience of natural ecosystems and their ability to persist in our fractured world[13]. A comprehensive collection of biodiversity data over a long time horizon has been used to explore the catalytic effects of landscape-scale forest loss on biodiversity change[14] and temperature-related biodiversity change[15]. Current literature is insufficient to explore biodiversity change. PAD events may reduce the systematic distribution of biodiversity in nearby areas, thus damaging the local ecological environment. Few studies have used data from protected areas. Past study has shown that protected area downsizing may exacerbate habitat fragmentation[16] and is a key contributor to biodiversity loss globally. No systematic research has been done on the impact of PAD events on biodiversity loss. Previous literature generally focused on quantifying the extent of PAD events, suggesting the habitat impact of PAD events[17,18]. This paper directly estimates the relationship between PAD events and biodiversity change by merging PAD events with biodiversity abundance data over a long time horizon.

## Results

### Abundance changes vary with distance

This study applies a two-way fixed effects model (or a generalized DID model) to estimate the relationship between PAD events and biodiversity changes. Results between abundance and the distance of proximate PAD events are in Supplementary Table 1. We find that abundance decreases after the nearby PAD events are enacted. Specifically, PAD events decrease the abundance of organisms by 29.7%. As the distance between records of abundance and PAD events increases, the effect of PAD events on abundance decreases. To test for the robustness of our result, we replace the dependent variables with the abundance measurement without taking the natural log and the natural logarithm of the biomass of species in sample. We find that the results are similar and the detailed regression coefficients replacing the dependent variables are displayed in Supplementary Tables 2 and 3.

In our model, the treatment group consists of abundance records with proximate PAD events within a certain range, while the control group includes those without proximate PAD events. A binned model that connects the proximate PAD events to abundance uncovers a more detailed relationship between PAD and biodiversity. We use 5 distance bins with an increment of 20-40 km to form a balanced distribution of the number of studies in each distance bin. We find that the proximity to the PAD event has decreased the abundance of biodiversity. The magnitudes of the abundance vary across different distance bins. Figure 2a plots the abundance changes caused by PAD events after controlling for meteorological variables, area, and yearly fixed effects. Results indicate a 28.1%, 24.2%, and 8.8% reduction in the biodiversity abundance within 0-20 km, 20-50 km, and 50-70 km, respectively. We observe that the negative impacts on the biodiversity abundance diminished with increasing distance from 0-70 km and the effect becomes insignificant beyond 70 km. Figure 2b indicates an abundance reduction of up to 26.0% when meteorological variables, yearly, and study fixed effects are included. The estimated coefficient becomes slightly smaller in magnitude when meteorological control variables are excluded (Supplementary Table 4). The smallest abundance reduction in magnitude is 21.7% for proximate PAD events within 20–50 km. In Figure 2b, we find the negative impact on biodiversity abundance ranges from 20 to 50 km. The impact becomes insignificant after the 50 km. Our results suggest that biodiversity is negatively affected by the proximity to PAD events within a 50 km buffer in general. We also use the abundance measurement without taking the natural log as the dependent variable. The detailed regression coefficients replacing the dependent variables are displayed in Supplementary Table 5. Our main results remain unchanged.

### Effects of PAD on nearby abundance

In the quasi-experimental DID model, we first determine the distance that separates the abundance data into control and treatment groups. We followed the standard methodology in the literature[19,20] and used the least local linear polynomial estimators to determine the distance from a PAD event at which the PAD event stops having a significant impact on the nearby abundance.

This cutoff distance was obtained from the intersection of the lower and upper confidence intervals of the residuals of the natural log of abundance versus the distance from the nearest PAD event that was formed before the nearest PAD event and the abundance that was formed after the nearest PAD event (Supplementary Figure 1). The residuals of the natural log of abundance were determined using an OLS model predicting the natural log of abundance versus the meteorological characteristics, study, and yearly fixed effects. Based on Supplementary Figure 1, the distance that separates the treatment and control groups is 110 km, suggesting the potential impact range of PAD events on biodiversity.

The results of the DID models are presented in Table 1. Model (1) includes area, and yearly fixed effects, as well as the set of control variables such as temperature and precipitations. Model (2) excludes the meteorological control variables from Model (1) and replaces the area fixed effects with the study fixed effects. Model (3) retains the same specifications as Model (2) with the inclusion of the control variables. Our focus is the coefficient associated with $post_{it} * treat_{it}$. We find that the estimates are statistically significant in Models (1) and (2) at a 1% level and a 10% level, respectively. In Model (3), we find that the $post_{it} * treat_{it}$ is still negative and statistically significant at a level close to 5% ($p$=0.054). Model (1) leads to a larger treatment effect estimate. Our preferred specification in Model (3) suggests an overall 32.3% decrease in abundance after the nearest PAD event is enacted. We also detect a significant negative impact when we replace the dependent variables with the natural logarithm of the biomass of species in sample to measure biodiversity based on the coefficient of $post_{it} * treat_{it}$ in Supplementary Table 6.

We conduct additional robustness checks to enhance the credibility of our main results. The analysis based on the event study model supports the parallel trend assumption. In the event study, we obtain the coefficients before and after the PAD events (Figure 3). We find that before the PAD events, the effects of PAD events are not statistically different from zero, which is consistent with the parallel trend assumption. After the PAD events, there starts to show negative impacts although the magnitude fluctuates, which supports our main results by ruling out the potential influences of differential trends.

**Heterogenous analyses based on species and PAD characteristics**
The abundance changes due to PAD events are heterogeneous across species and PAD characteristics. To investigate potential mechanisms of the PAD event impact, we included additional interaction terms where the $post_{it} * treat_{it}$ is interacted with the species and PAD characteristic variables in Figure 4 and Supplementary Table 7. We examine how abundance changes over several key species characteristics, including the realm of site, habitat, biome as listed on the WWF (World Wildlife Foundation) site, and taxa. Our results reveal that biodiversity loss near water is more severe in response to PAD events, based on the comparison of organisms in marine and freshwater with those in terrestrial. We also observe positive significant effects on abundance changes caused by organisms in the terrestrial realm, habitats such as grasslands and forests, and biome like shrublands. The negative impacts on abundance are mostly attributed to organisms living in contact with water. PAD events decrease the abundance of organisms living in contact with water by over 70%, but increase the abundance of terrestrial organisms by around 30%. PAD events reduce restrictions on overfishing, pollution, invasive species, underwater noise, development activities, climate change, ocean acidification, and others, which have a greater impact on the living environment of marine and freshwater life. We also find PAD events have a negative impact on the abundance in non-mammals while a positive impact on those in mammals. Specifically, enacting PAD events significantly reduces abundance in non-mammals by 47.2%.

We also examine how abundance variations change if the PAD event is later reversed and when the species belongs to the IUCN category. We find that PAD events have significantly negative impacts on organisms close to reversed PAD events and organisms near PA belonging to the IUCN category. Enacting PAD events significantly reduces abundance by 36.3% in areas where PAD events were later reversed as well as in areas that were in the IUCN category before PAD. Possible explanations include policy reversal reduces management effectiveness in the short term, the IUCN management category cannot inhibit the aggravation of human pressure caused by PAD events[1]. The areas that receive more attention from the government are vulnerable to events that trigger biodiversity loss.

**Conclusion**

This study examines how PAD events affect biodiversity. We find that PAD events significantly reduce abundance based on micro-level data. The proximity to PAD events affects biodiversity negatively and decreases the abundance by 26.0% within 50 km. We also observe an overall 32.3% decrease in abundance after the nearest PAD event is enacted. We find that the negative impacts on biodiversity originate from organisms living in contact with water, non-mammals, organisms close to reversed PAD events, and organisms near PA belonging to the IUCN category.

Existing estimates of the mean willingness to pay for biodiversity ranging from $0 to $6.39 per resident[21]. Our estimations imply that the direct economic value in abundance loss due to the nearest PAD event is approximately up to $2.06 per resident per year. National losses add up to $689.95 million in 2022 (see Methods). The large economic loss highlights the importance of addressing potential environmental problems caused by PAD events. This study also calls for the establishment of different forms of nature reserves to protect biodiversity, such as national parks. Adopting such measures is helpful to increase public willingness to pay for biodiversity conservation and other environmental protection initiatives.

Many nations have aggressive expansion plans for energy, natural resources, transportation industries, and related infrastructures[22]. PAD events encourage economic exploitation while facilitating development objectives. The cost of economic activities may be lower in lands that used to be PAs than in other regions. However, damaged ecosystems may hinder sustainable economic development. The degradation of PAs may also influence PAs and have long-term effects on ecosystems[4]. The loss of biodiversity due to PAD events may lead to an irreversibly degraded ecosystem. Policymakers should consider the potential loss of biodiversity in the cost-benefit analysis of PAD events. Moreover, governments in other countries with some PAD events, such as the United Kingdom, Australia, Brazil, and South Africa, can quantify the impacts of PAD events on biodiversity and design more practical strategies for PA development.

PAD events can also erode biodiversity at a regional level. It is important to set plans and policies aligned with a long-term regional vision for biodiversity conservation to minimize infrastructure expansion, halt extensive biodiversity losses surrounding roads, and manage the landscape-wide consequences. Strategic regional plans, and other conservation policies governing the effects of land-use change like management of the surviving PA network, can be used to control threats to biodiversity and mitigate the cascading effects on organisms, ecosystems, and PAs[4].

There are several areas for future research. Due to data limitation, this study does not consider PADDD events primarily in marine systems and on private lands[8]. While we provide estimates of PAD events on abundance, the impacts of PAD events on biodiversity may be more accurate when integrating PADDD tracking data with other area-based conservation databases (such as the World Database on PAs). Considering the critical role of biodiversity in the ecosystem, further research on estimating the impact of PAD events on biodiversity, quantifying the under-appreciated cost-saving benefits of effective biodiversity conservation and the corresponding economic loss in the ecosystem will be valuable. For instance, it would be helpful to calculate the costs of the socioeconomic destruction caused by zoonotic diseases against those of managing PAs which lessens supply to illegal wildlife markets.

**Method**
**Data sources**
We obtained biodiversity data from BioTIME (BioTIME— Global database of biodiversity time series (st-andrews.ac.uk)), the largest database of assemblage time series currently available. BioTIME contains hundreds of academic ecological research studies measuring the abundance (count or biomass) of pertinent species in a given area over time, spanning several decades[23,24]. BioTIME is composed of species abundance records for assemblages that have been sampled through time using a consistent methodology. The dataset included 381 separate studies (study ID plus additional data sources) that covered a variety of taxonomic groups, including plants, invertebrates, birds, mammals, and fish. As a result, the BioTIME data is useful for investigating the relationship between economic development and biodiversity based on the rich set of features. For our analysis, we used studies from the continental US terrestrial system between 1903 and 2018. Since there aren't enough freshwater and marine studies, we didn't include these two categories in our analysis of biodiversity trends across taxa and geographic regions. The website Padddtracker.org provided detailed information on PAD events, which chronicled legal changes for PAD events, collated available data on a worldwide scale using records that had already been published as well as those that had not, and updated versions created by the World Wildlife Fund and Conservation International. The dataset includes information about the latitude-longitude location of all known PAD events.

Our analysis focuses on PAD events. We mapped all the PADDD and biodiversity information on a global scale using the BioTIME and PADDDtracker data. Then we adjusted our dataset to focus on PAD events located within the continental U.S. (Figure 1). To implement this adjustment, a GIS shapefile was imported for the continental U.S. and we matched PAD events to find the PAs that are located within the continental U.S.. Information about the biodiversity of protected areas at each PAD event across the continental U.S. was determined using ArcGIS by linking the biodiversity data to the closest PAs. The distance between biodiversity data and the PA is also calculated. Furthermore, Stata is used to merge biodiversity data and information on PAD events with identical STUDY_ID. As a result, all the PAD events closest to the biodiversity data are included in our sample.

To compare regions with different climate types and to control for environmental conditions, we added average annual minimum temperature, extremely high temperature (highest daily maximum temperature for the year), and total annual precipitation as covariables. The meteorological data were retrieved from the Global Summary of the Year dataset provided by the National Oceanic and Atmospheric Administration (NOAA)/National Centers for Environmental Information, computed using the Global Historical Climatology Network (GHCN)-Daily Data set. We merged meteorological control variables with the main data set by using latitude-longitude location. A list of summary statistics for the PAD events and related variables is provided in Supplementary Table 8.

**Bin approach**
A two-way fixed effects model (or generalized DID model) enables us to take advantage of the panel data structure. The two-way fixed effects model is specified in equation (1) below:

$$\mathrm{Log}\, Y_{it} = \beta_0 + \beta_1 post_{it} + \beta_2 post_{it} * distance_{it} + \gamma X_{it} + \mu_t + \theta_p + \delta_s + \epsilon, \qquad (1)$$

where $Y_{it}$ is the abundance of species in sample *i* at year *t*. The natural logarithm form of the abundance is used as the explanatory variable. The species abundance records are from assemblages consistently sampled for a minimum of 2 years. About 81% of the samples were observed for at least ten consecutive years. $post_{it}$ is a time dummy variable to indicate whether the abundance record occurs before or after the closest PAD event, and $post_{it}$ takes 1 if the abundance record occurs after the completion, and 0 otherwise. $distance_{it}$ represents the distance between records of abundance of species and the nearest PAD event. To better control for the differences in properties between the treatment and control groups, a series of meteorological control variables $X_{it}$ are added, including annual minimum temperature, extremely high temperature, and precipitation. Area fixed effects $\theta_p$ account for PAD-specific time-invariant factors; study fixed effects $\delta_s$ control for the perturbation of the abundance of species by unobservable factors that do not change over time; yearly fixed effects $\mu_t$ control for time-invariant characteristics. $\epsilon$ is an idiosyncratic error term.

Proximity to PAD events could make a big difference, and area fixed effects can capture these structural differences so that we can focus on the comparison of biodiversity near similar PAD events and avoid comparing biological habitats of different characteristics. The binned model uncovers a distanced-based relationship between PAD and biodiversity, which is specified in equation (2):

$$\text{Log } Y_{it} = \beta_0 + \beta_1 post_{it} + \beta_{2,b} post_{itb} * vicinity_{itb} + \gamma X_{it} + \mu_t + \theta_p + \delta_s + \epsilon. \quad (2)$$

The variable $vicinity_{itb}$ indicates the proximity of the abundance record to the closest PAD event, with $vicinity_{itb}$ equals 1 if an abundance record is close to a PAD event within a distance bin $b$, and 0 otherwise. We applied 5 distance bins based on an increment of 20-40 km, with the distance higher than 110 km included in the last bin. $\beta_{2,b}$ is the coefficient for the interaction term between $vicinity_{itb}$ and $post_{it}$, and represents the impact difference between the treatment and control groups on biodiversity. $\beta_{2,b}$ identifies the treatment effect changes as the vicinity increases after the PAD.

**Treatment and Control (DID, quasi-experiment approach)**
We also run a quasi-experimental DID analysis to provide more evidence on the choice of the cutoff distance. This analysis confirms that overall significant changes happen within 110 km. Let $Log\ Y_{it}$ be the natural logarithm of the abundance of species in sample $i$ at year $t$. The treatment group is defined as those that are close to PAD event enough (i.e., closer than 110 kilometers) to be affected. The dummy variable $treat_i$ is equal to 1 if abundance record $i$ belongs to the treatment group (i.e., is located surrounding PAD event less than 110 kilometers ), and equal 0 if it belongs to the control group (i.e., outside 110 kilometers). Let $post_{it}$ take 1 if the abundance record occurs after the completion, and 0 otherwise. The DID model can be written as equation (3):

$$\text{Log } Y_{it} = \beta_0 + \beta_1 post_{it} + \beta_2 treat_i + \pi post_{it} * treat_i + \gamma X_{it} + \mu_t + \theta_p + \delta_s + \epsilon \quad (3)$$

where $\pi$ represents the treatment effect of the PAD events on the biodiversity changes by comparing the differences between the treatment and control groups before and after the PAD events.

**Event study**

To test the plausibility of the parallel trend assumption between the records of abundance with proximate PAD events and those without, we conducted an event study analysis. The event study model is specified as follows:

$$\text{Log} Y_{it} = \beta_0 + \sum_{j=-10}^{j=10} \pi_j [Treat_i \times I(t - T_i = j)] + \gamma X_{it} + \mu_t + \theta_p + \delta_s + \epsilon \quad (4)$$

where $Treat_i$ is a dummy variable that takes the value of one if abundance record $i$ belongs to the treatment group and takes the value of zero otherwise belongs to the control group, which is the same as in equation (3). $T_i$ is the specific year $t$ when abundance record $i$ enacted a PAD event. $I(\cdot)$ is an indicator that equals one when $(t - T_i = j)$ and zero otherwise. The baseline, omitted case is the two years before the PAD event was enacted ($j$ = -2 to 0). All other variables carry the same definitions as in equation (1). The coefficients $\pi_j$ measure the effects of PAD events on biodiversity in the relative year $j$, compared to that in two years before the PAD event was enacted ($j$ = -2 to 0). If $\pi_{-10}$ to $\pi_{-3}$ are not statistically significant, the abundance in the two groups is statistically indifferent before the PAD events, suggesting the plausibility of the parallel trend assumption.

**Direct Economic Loss Assessment by a Back-To-Envelope Model**

Biodiversity generates both direct and indirect economic value in terms of how changes in biodiversity affect human well-being[25]. Direct economic value is derived from production and consumption or interaction with environmental resources and services[26]. The willingness to pay (WTP) for conserving a particular species measures the direct economic value of biodiversity[27]. The indirect economic value of biodiversity relates to the indirect support for the ecosystem's stability and survival and protection provided to economic activity and property by the ecosystem's natural functions, or regulatory environmental services[28], which is challenging to quantify[29]. In this study, we focus on the direct economic loss of biodiversity and collect the WTP data from the literature.[28] Using a back-of-the-envelope estimation, we calculate the annual direct economic loss in biodiversity caused by the closest PAD event:

$$Loss = (\pi \times 100\%) \times \overline{WTP} \quad (7)$$

where $Loss$ is the mean economic loss per resident per year due to a change in abundance caused by the nearest PAD event. $\pi$ is the estimated treatment effect of the nearest PAD event on abundance based on equation (3) by a DID estimation. $\overline{WTP}$ is the mean willingness to pay for biodiversity per resident per year, which ranges from $0 to $6.39[30] and is equivalent to a $2.06

loss in abundance per resident per year from the nearest PAD event. The national losses in abundance thus add up to $2.06×334,282,669=$689.95$ million , where 334,282,669 is the population in the U.S. in 2022.

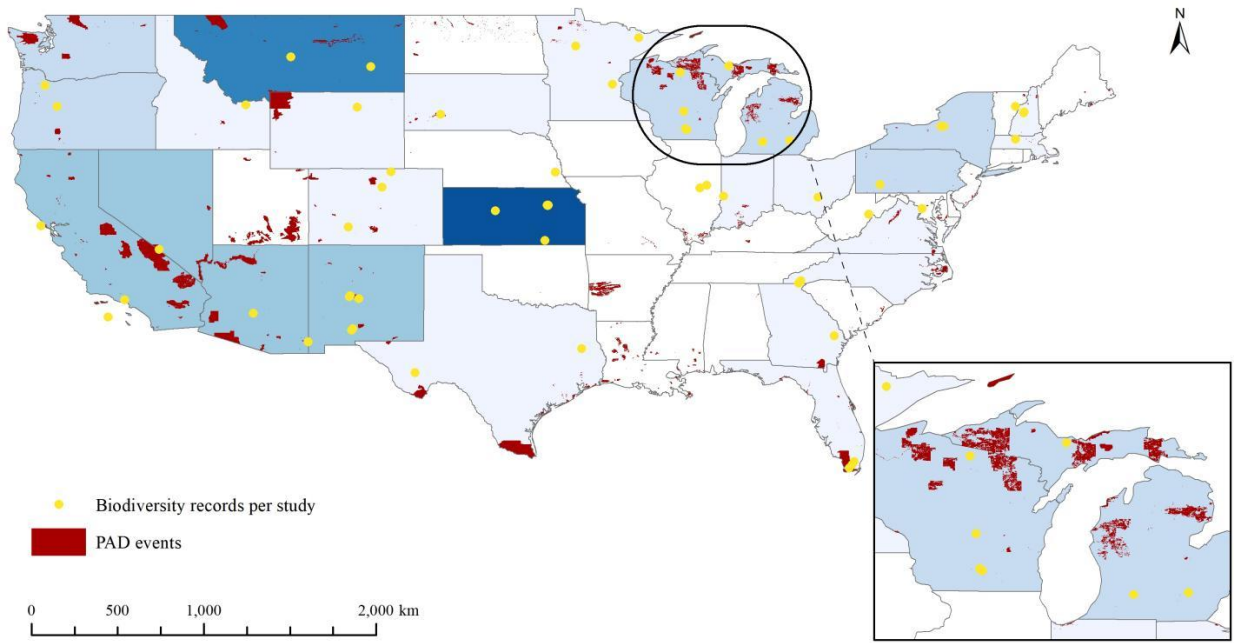

**Figure 1 Spatial distribution of sample PAD events and biodiversity in the continental U.S. from 1903 to 2018.** Dots indicate biodiversity records per study in BioTIME, which holds millions of records of species counts at the species-location (latitude and longitude)-year level at more than 10,000 different locations. Red shaded areas show enacted PAD events. The blue background represents the states with biodiversity records affected by adjacent PAD events, with darker blue colors indicating a larger number of specific biodiversity records.

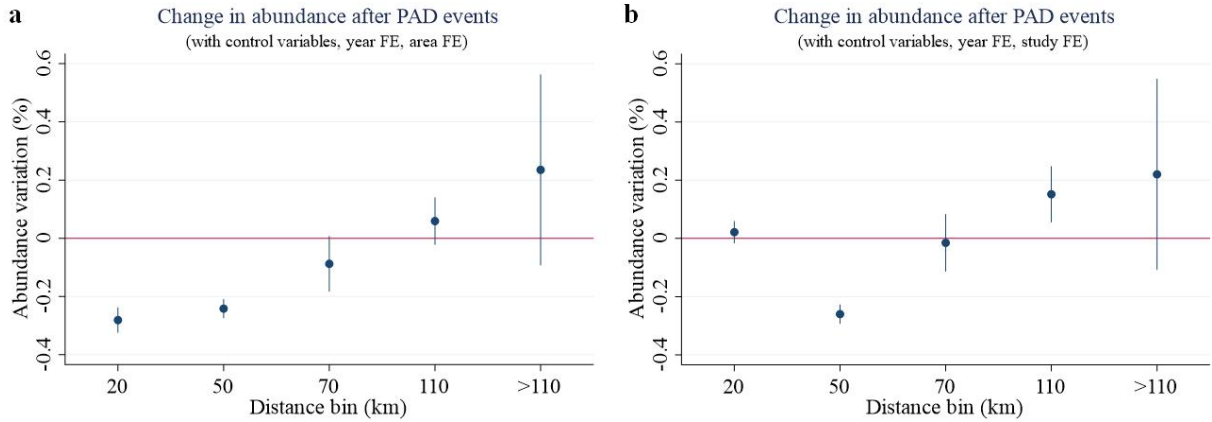

**Figure 2 Impacts of vicinity to PAD events on abundance. a**, Impacts with meteorological variables, year, and area fixed effects. **b**, Impacts with meteorological variables, year, and study fixed effects. The centers of the error bars are the values of the coefficients, which represent point estimates from the regressions and indicate the average effects of the PAD events. Their 95% confidence intervals are plotted vertically. The dependent variable is the log of species abundance. Each panel is from one regression. For both regressions, the total number of observations is over 1.9 million each.

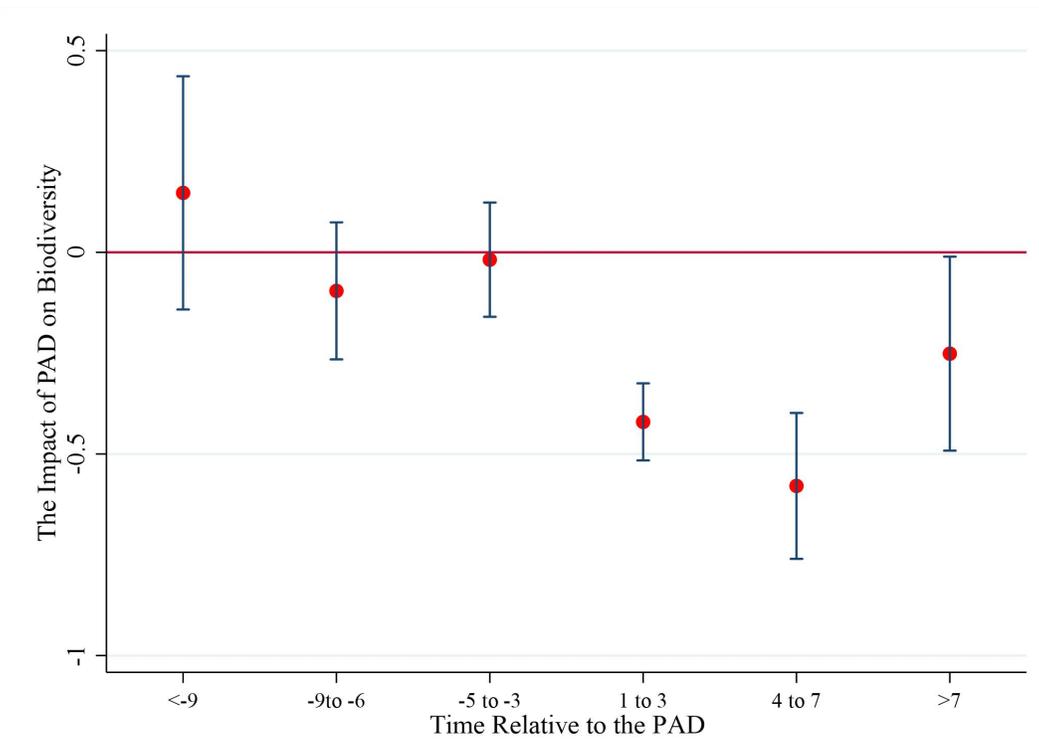

**Figure 3 Abundance changes based on the dynamic event study model.** This figure provides the abundance changes for biodiversity relative to the year of the PAD event. The red solid line represents the effect of the year of the PAD event. The horizontal axis is normalized relative to the year of the treatment and the excluded period is t=-2 to 0. The blue whiskers indicate the 95% confidence intervals of point estimates that show average effects. The year fixed effects, area fixed effects, and study fixed effects are included. The number of observations is 207,869. We have dropped the observations before t = −10 and after t = 10. The effects do not seem to be lasting and fade away after around 7 years.

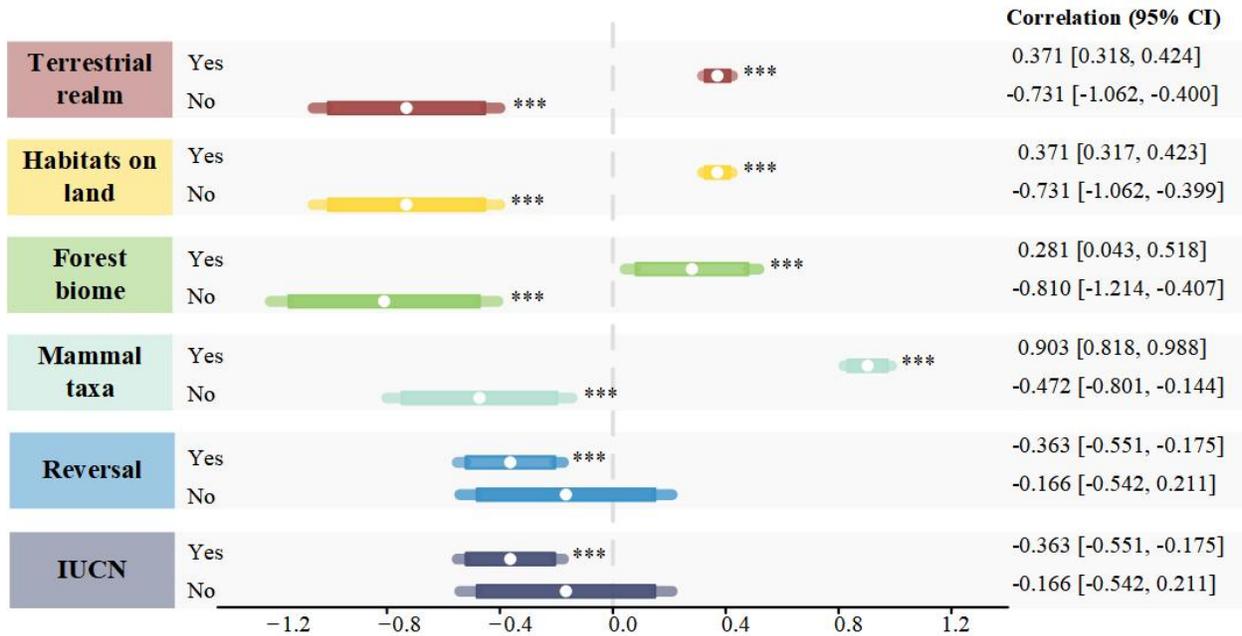

**Figure 4 Heterogenous effects across species and PAD characteristics.** Species characteristics include the realm of site (organisms in terrestrial and marine/freshwater), habitat (areas on land like forest/desert/grassland and areas near the water like lakes/ponds/streams), biome as listed on the WWF site (organisms far from water like shrublands/forest/grasslands and organisms living in contact with water like lake ecosystems/river ecosystems/shelf ecoregions), as well as taxa (mammal and non-mammal). As for PAD characteristics, reversal refers to a dummy of whether the species lives in areas where PAD events were later reversed, while IUCN is a dummy of whether the species is in areas that were listed in the IUCN category before PAD events. Estimates of the impact of the PAD events on the abundance are depicted by the white dots. The outer (thin) error bar and inner (thick) error bar for each estimate, respectively, indicate the 95% and 90% confidence intervals. The grey dashed line is at an estimated effect of zero. * $p<0.1$, ** $p<0.05$, *** $p<0.01$.

**Table 1 DID estimation results on abundance**

| Variables | (1) | (2) | (3) |
|---|---|---|---|
| Post×treat | -0.525*** | -0.287* | -0.323* |
|  | (0.167) | (0.168) | (0.168) |
| Post | 0.230 | 0.207 | 0.221 |
|  | (0.167) | (0.167) | (0.167) |
| Treat | -0.661*** |  |  |
|  | (0.008) |  |  |
| Constant | 2.007*** | 1.213*** | 1.400*** |
|  | (0.021) | (0.002) | (0.020) |
| Control Variables | YES |  | YES |
| Yearly FE | YES | YES | YES |
| Area FE | YES |  |  |
| Study FE |  | YES | YES |
| N | 1,902,298 | 1,902,629 | 1,902,298 |
| $R^2$ | 0.541 | 0.563 | 0.563 |

Notes: The dependent variable is the log of abundance. * $p<0.1$, ** $p<0.05$, *** $p<0.01$. Standard errors are in the parentheses.

# Supplementary Information

## Supplementary Figures

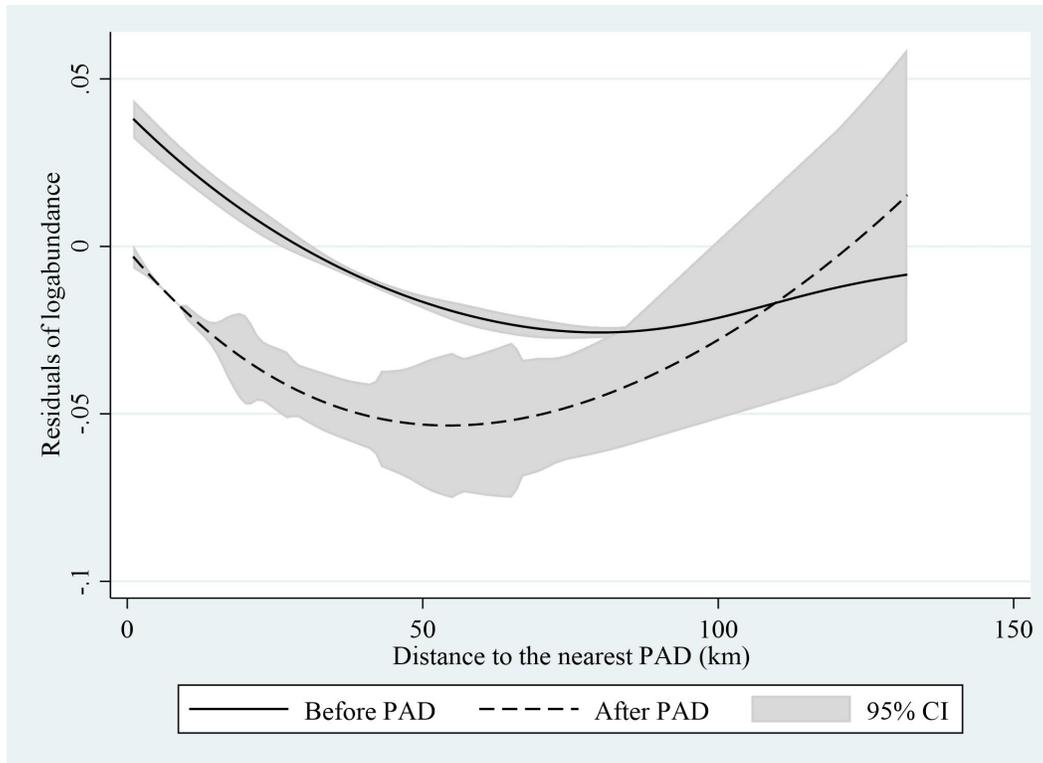

**Supplementary Figure 1 Abundance changes of distance from PAD events.** The abundance records are scattered across the country. As the distance increases, fewer abundance records are affected by enacting PAD events, and the confidence intervals get wider. The solid and dashed lines represent the residuals of the natural log of abundance that was formed before the nearest PAD event and was formed after the nearest PAD event, respectively. The shaded areas depict the 95% confidence intervals.

# Supplementary Tables

**Supplementary Table 1 The impacts of the distance of proximate PAD events on abundance**

| Variables | (1) | (2) | (3) |
|---|---|---|---|
| Post | -0.297*** | -0.085*** | -0.080*** |
|  | (0.026) | (0.025) | (0.025) |
| Post×distance | 0.002** | 0.001 | -0.001 |
|  | (0.001) | (0.001) | (0.001) |
| Constant | 1.593*** | 1.214*** | 1.398*** |
|  | (0.021) | (0.003) | (0.020) |
| Control Variables | YES |  | YES |
| Yearly FE | YES | YES | YES |
| Area FE | YES |  |  |
| Study FE |  | YES | YES |
| N | 1,902,298 | 1,902,629 | 1,902,298 |
| $R^2$ | 0.539 | 0.563 | 0.563 |

Notes: The dependent variable is the log of abundance. * $p<0.1$, ** $p<0.05$, *** $p<0.01$. Standard errors are in the parentheses.

**Supplementary Table 2 The impacts of the distance of proximate PAD events are robust to abundance**

| Variables | (1) | (2) | (3) |
|---|---|---|---|
| Post | -6407.506*** | -4660.212*** | -4693.811*** |
|  | (729.105) | (720.335) | (720.234) |
| Post×distance | 66.352*** | 72.592*** | 73.390*** |
|  | (14.213) | (14.608) | (14.825) |
| Constant | 792.516 | 2064.349*** | 568.421 |
|  | (703.785) | (112.223) | (719.194) |
| Control Variables | YES |  | YES |
| Yearly FE | YES | YES | YES |
| Area FE | YES |  |  |
| Study FE |  | YES | YES |
| N | 1,902,298 | 1,902,629 | 1,902,298 |
| $R^2$ | 0.071 | 0.074 | 0.074 |

Notes: The dependent variable is the abundance measurement without taking the natural log. * $p<0.1$, ** $p<0.05$, *** $p<0.01$. Standard errors are in the parentheses.

**Supplementary Table 3 The impacts of the distance of proximate PAD events are robust to biomass**

| Variables | (1) | (2) | (3) |
|---|---|---|---|
| Post | 0.019 | -0.185*** | -0.198*** |
|  | (0.068) | (0.067) | (0.068) |
| Post×distance | 0.007*** | 0.008*** | 0.008*** |
|  | (0.002) | (0.002) | (0.002) |
| Constant | 0.390*** | 1.130*** | 0.651*** |
|  | (0.059) | (0.002) | (0.064) |
| Control Variables | YES |  | YES |
| Yearly FE | YES | YES | YES |
| Area FE | YES |  |  |
| Study FE |  | YES | YES |
| N | 888,753 | 888,753 | 888,753 |
| $R^2$ | 0.352 | 0.419 | 0.419 |

Notes: The dependent variable is the natural logarithm of the biomass. * $p<0.1$, ** $p<0.05$, *** $p<0.01$. Standard errors are in the parentheses.

**Supplementary Table 4 The impacts of proximity to PAD events on abundance**

| Variables | (1) | (2) | (3) |
|---|---|---|---|
| Vicinity 0-20 km | -0.281*** | 0.029 | 0.022 |
|  | (0.022) | (0.019) | (0.019) |
| Vicinity 20-50 km | -0.242*** | -0.217*** | -0.260*** |
|  | (0.016) | (0.016) | (0.017) |
| Vicinity 50-70 km | -0.088* | 0.012 | -0.015 |
|  | (0.048) | (0.050) | (0.050) |
| Vicinity 70-110 km | 0.059 | 0.128*** | 0.152*** |
|  | (0.041) | (0.049) | (0.049) |
| Vicinity >110 km | 0.235 | 0.206 | 0.220 |
|  | (0.167) | (0.167) | (0.167) |
| Constant | 1.592*** | 1.197*** | 1.414*** |
|  | (0.021) | (0.003) | (0.020) |
| Control Variables | YES |  | YES |
| Yearly FE | YES | YES | YES |
| Area FE | YES |  |  |
| Study FE |  | YES | YES |
| N | 1,902,298 | 1,902,629 | 1,902,298 |
| $R^2$ | 0.539 | 0.563 | 0.563 |

Notes: The dependent variable is the log of abundance. These vicinity coefficients are interacted with the dummy variable of post PAD events. * $p<0.1$, ** $p<0.05$, *** $p<0.01$. Standard errors are in the parentheses.

**Supplementary Table 5 The impacts of proximity to PAD events are robust to abundance**

| Variables | (1) | (2) | (3) |
|---|---|---|---|
| Vicinity 0-20 km | -7603.813*** | -4045.751*** | -4119.723*** |
|  | (738.319) | (691.344) | (692.456) |
| Vicinity 20-50 km | -1785.272*** | -1790.928*** | -1721.240*** |
|  | (188.913) | (166.343) | (198.620) |
| Vicinity 50-70 km | -2809.603*** | -2433.665*** | -2425.220*** |
|  | (282.199) | (264.404) | (281.623) |
| Vicinity 70-110 km | -1609.983*** | -802.030*** | -997.392*** |
|  | (121.946) | (94.656) | (117.549) |
| Vicinity >110 km | -126.496** | -229.936*** | -169.484*** |
|  | (63.686) | (50.413) | (64.484) |
| Constant | 414.049 | 2018.790*** | 370.803 |
|  | (710.072) | (115.086) | (725.319) |
| Control Variables | YES |  | YES |
| Yearly FE | YES | YES | YES |
| Area FE | YES |  |  |
| Study FE |  | YES | YES |
| N | 1,902,298 | 1,902,629 | 1,902,298 |
| $R^2$ | 0.071 | 0.074 | 0.074 |

Notes: The dependent variable is the abundance measurement without taking the natural log. These vicinity coefficients are interacted with the dummy variable of post PAD events. * $p<0.1$, ** $p<0.05$, *** $p<0.01$. Standard errors are in the parentheses.

**Supplementary Table 6 DID estimation results are robust to biomass**

| Variables | (1) | (2) | (3) |
|---|---|---|---|
| Post×treat | -0.460*** | -0.513*** | -0.526*** |
|  | (0.112) | (0.111) | (0.111) |
| Post | 0.608*** | 0.472*** | 0.476*** |
|  | (0.101) | (0.100) | (0.100) |
| Constant | 0.390*** | 1.130*** | 0.651*** |
|  | (0.059) | (0.002) | (0.064) |
| Control Variables | YES |  | YES |
| Yearly FE | YES | YES | YES |
| Area FE | YES |  |  |
| Study FE |  | YES | YES |
| N | 888,753 | 888,753 | 888,753 |
| $R^2$ | 0.352 | 0.419 | 0.419 |

Notes: The dependent variable is the natural logarithm of the biomass. * $p<0.1$, ** $p<0.05$, *** $p<0.01$. Standard errors are in the parentheses.

**Supplementary Table 7 Heterogeneous effects of PAD events on abundance**

| Variables | (1) | (2) | (3) | (4) | (5) | (6) |
|---|---|---|---|---|---|---|
| Post×treat | -0.731*** | -0.731*** | -0.810*** | -0.472*** | -0.166 | -0.166 |
|  | (0.169) | (0.169) | (0.206) | (0.168) | (0.192) | (0.192) |
| Post×treat×terrestrial realm | 0.371*** |  |  |  |  |  |
|  | (0.027) |  |  |  |  |  |
| Post×treat×habitats on land |  | 0.371*** |  |  |  |  |
|  |  | (0.027) |  |  |  |  |
| Post×treat×forest biome |  |  | 0.281** |  |  |  |
|  |  |  | (0.121) |  |  |  |
| Post×treat×mammal taxa |  |  |  | 0.903*** |  |  |
|  |  |  |  | (0.043) |  |  |
| Post×treat×reversal |  |  |  |  | -0.363*** |  |
|  |  |  |  |  | (0.096) |  |
| Post×treat×IUCN |  |  |  |  |  | -0.363*** |
|  |  |  |  |  |  | (0.096) |
| Constant | 0.952*** | 0.897*** | 3.807*** | 1.921*** | 1.955*** | 1.973*** |
|  | (0.063) | (0.068) | (0.079) | (0.021) | (0.025) | (0.023) |
| Control Variables | YES | YES | YES | YES | YES | YES |
| Yearly FE | YES | YES | YES | YES | YES | YES |
| Area FE | YES | YES | YES | YES | YES | YES |
| N | 1902298 | 1902298 | 1902298 | 1902298 | 1902298 | 1902298 |
| $R^2$ | 0.542 | 0.542 | 0.541 | 0.543 | 0.541 | 0.541 |

Notes: We only presented the coefficient estimates for the $post_{it} * treat_{it}$ variable and its associated interaction terms with the species and PAD characteristics. * $p<0.1$, ** $p<0.05$, *** $p<0.01$. Standard errors are in the parentheses.

**Supplementary Table 8 Summary statistics**

| Variables | N | Mean | SD | Min | Max |
|---|---|---|---|---|---|
| Year | 2,660,148 | 1,995 | 15.48 | 1,903 | 2,018 |
| Yearpost | 2,660,148 | -14.04 | 24.97 | -113 | 81 |
| Post | 2,660,148 | 0.120 | 0.325 | 0 | 1 |
| Abundance | 1,902,629 | 1,368 | 49,622 | 0.000900 | 37,500,000 |
| Biomass | 888,753 | 44.80 | 390.2 | 0.000004 | 18,171 |
| Logabundance | 1,902,629 | 1.200 | 1.661 | -7.013 | 17.44 |
| Logbiomass | 888,753 | 1.129 | 2.094 | -12.43 | 9.808 |
| Post*distance | 2,660,148 | 0.867 | 5.298 | 0 | 131.9 |
| Extremely high temperature | 2,660,148 | 36.28 | 4.073 | -2.500 | 64.10 |
| Minimum temperature | 2,660,148 | 5.352 | 4.573 | -14.68 | 25.32 |
| Precipitation | 2,660,148 | 856.2 | 306.1 | 11.50 | 2,852 |
| Vicinity 0-20 km | 2,660,148 | 0.183 | 0.386 | 0 | 1 |
| Vicinity 20-30 km | 2,660,148 | 0.0220 | 0.147 | 0 | 1 |
| Vicinity 30-50 km | 2,660,148 | 0.0652 | 0.247 | 0 | 1 |
| Vicinity 50-75 km | 2,660,148 | 0.249 | 0.433 | 0 | 1 |
| Vicinity >75 km | 2,660,148 | 0.481 | 0.500 | 0 | 1 |
| Realm | 2,660,148 | 0.831 | 0.374 | 0 | 1 |
| Habitat | 2,660,148 | 0.847 | 0.360 | 0 | 1 |
| Biome | 2,660,148 | 0.942 | 0.233 | 0 | 1 |
| Taxa | 2,660,148 | 0.0272 | 0.163 | 0 | 1 |
| Reversal | 2,660,148 | 0.306 | 0.461 | 0 | 1 |
| IUCN | 2,660,148 | 0.930 | 0.255 | 0 | 1 |